# Hidden Time-Reversal Symmetry in Dissipative Reciprocal Systems

MÁRIO G. SILVEIRINHA,[1*]

[1] *University of Lisbon – Instituto Superior Técnico and Instituto de Telecomunicações, Avenida Rovisco Pais 1, 1049-001 Lisboa, Portugal*
*mario.silveirinha@co.it.pt*

**Abstract:** It is proven, without using the microscopic reversibility argument of Onsager, that lossy reciprocal systems have a hidden time-reversal symmetry. The key idea is that the dissipation channels of lossy dielectrics can be mimicked by a distributed network of lossless transmission lines. It is highlighted that the reciprocity of lossy dielectrics is fundamentally rooted on the hidden time-reversal invariance and on linearity of the materials. Furthermore, it is demonstrated that the upper-half plane response of dissipative materials can be approximated as much as desired by the response of some lossless material.

## 1. Introduction

The laws that rule the propagation of light in free space are invariant under the "time reversal" operation [1, 2, 3]. This means that when the arrow of time is flipped – so that the dynamics of the electromagnetic wave is reversed in time similar to a movie played backwards –the time-reversed wave remains compatible with the laws of electromagnetism, i.e., it still satisfies the Maxwell equations.

An extraordinary consequence of the time-reversal invariance is that the propagation of light in typical material platforms is inherently bi-directional. For example, if a wave can go through some channel with no back-reflections, then the time-reversed wave can go through the same channel but propagating in the opposite direction. This rather profound and intriguing bi-directional character of photonic systems is usually regarded a consequence of the Lorentz reciprocity law [4-5], which is more general than the time-reversal invariance. Indeed, the Lorentz reciprocity theorem can be applied to *macroscopic* systems with dissipative elements, which are not invariant under a time-reversal transformation.

The reciprocity of dissipative systems is usually justified by the microscopic reversibility of physical processes [2, 6], i.e., relying on the time-reversal invariance of (classical) physics at the *microscopic* level (the quantum world is also mostly ruled by time-reversal invariant laws, with a few exceptions related to nuclear physics which are unimportant for the effects discussed here). Microscopic reversibility and statistical mechanics were used by Onsager to derive reciprocal relations for irreversible processes [2, 6].

In this article, I show that it is unnecessary to invoke microscopic arguments to establish a direct link between the reciprocity of macroscopic systems and time-reversal invariance. To this end, I prove that dissipative dielectrics have a hidden time-reversal symmetry, as the relevant dissipation channels can be mimicked by a distributed network of infinitely-extended time-reversal invariant lossless transmission lines. Using this result, I prove that the Lorentz reciprocity is ultimately a consequence of the hidden time-reversal invariance, linearity and conservation of energy. Furthermore, I also show that the response of a passive reciprocal dissipative material can be approximated as much as desired (in a given time interval with finite duration) by that of a time-reversal invariant lossless medium.

Previous studies of quantum electrodynamics in lossy material platforms [7-12] have revealed a rather profound link between dissipation, "open systems", and interactions with a "bath" of oscillators. For example, Ref. [11] demonstrates that a lossy electric circuit described by some impedance function can be implemented using lossless networks of idealized inductors and capacitors. Conversely, the dynamics of some classical inhomogeneous systems is determined by a "bath" of harmonic oscillators, and due to this reason their response is effectively dissipative, even when they are formed exclusively by lossless materials [13, 14]. This article exploits and extends these known paradigms to unveil the hidden-symmetry of reciprocal materials.

## 2. Time-reversal symmetry

The propagation of light in composite media is ruled by the macroscopic Maxwell equations:

$$\nabla \times \mathbf{E} = -\frac{\partial \mathbf{B}}{\partial t}, \qquad \nabla \times \mathbf{H} = \mathbf{j} + \frac{\partial \mathbf{D}}{\partial t}. \tag{1}$$

The time-reversal operation $\mathcal{T}$ transforms the electromagnetic fields $\mathbf{E}, \mathbf{H}$ as [2]:

$$\mathbf{E}(\mathbf{r},t) \xrightarrow{\mathcal{T}} \mathbf{E}^{TR} = \mathbf{E}(\mathbf{r},-t), \qquad \mathbf{H}(\mathbf{r},t) \xrightarrow{\mathcal{T}} \mathbf{H}^{TR} = -\mathbf{H}(\mathbf{r},-t) \tag{2}$$

The electric displacement vector $\mathbf{D}$ is transformed similarly to $\mathbf{E}$, and the induction field $\mathbf{B}$ and the electric current density $\mathbf{j}$ in the same manner as $\mathbf{H}$. Thus, the former is said to be *even* under the time reversal operation, whereas the latter are *odd*. Similar to the original fields, the time-reversed fields $\mathbf{E}^{TR}, \mathbf{H}^{TR}, \mathbf{j}^{TR},...$ satisfy the Maxwell's equations (1). A different (pseudo-) time-reversal symmetry for photonic systems was introduced in Ref. [15].

For a idealized lossless dielectric material with an instantaneous response the constitutive relations are $\mathbf{D} = \varepsilon(\mathbf{r})\mathbf{E}$ and $\mathbf{B} = \mu_0 \mathbf{H}$. Evidently, the time-reversed fields are linked in the same manner, i.e., $\mathbf{D}^{TR} = \varepsilon(\mathbf{r})\mathbf{E}^{TR}$ and $\mathbf{B}^{TR} = \mu_0 \mathbf{H}^{TR}$. Thereby, dielectric platforms formed by lossless dielectrics are time-reversal invariant. This property also holds true when the (lossless) materials are dispersive. Indeed, the electrodynamics of lossless materials can be formulated as a Schrödinger-type time evolution problem [16-19], which can be shown to be time-reversal invariant for dispersive isotropic dielectrics. Moreover, typical lossless nonlinear platforms are time-reversal invariant [20].

For linear media, the Maxwell equations admit time-harmonic solutions with a time variation $e^{-i\omega t}$, with $\omega$ the *real-valued* oscillation frequency. The time-dependent electromagnetic fields and current density are written in terms of complex amplitudes in the usual way, e.g., $\mathbf{E}(\mathbf{r},t) = \text{Re}\{\mathbf{E}_\omega(\mathbf{r})e^{-i\omega t}\}$. Evidently, $\mathcal{T}$ acts on the electric field as $\mathbf{E}(\mathbf{r},t) \to \text{Re}\{\mathbf{E}_\omega(\mathbf{r})e^{+i\omega t}\} = \text{Re}\{\mathbf{E}_\omega^*(\mathbf{r})e^{-i\omega t}\}$. Thus, in the frequency domain the time-reversal operation is closely related to a complex conjugation [21]. The complex field amplitudes are transformed as:

$$\mathbf{E}_\omega(\mathbf{r}) \xrightarrow{\mathcal{T}} \mathbf{E}_\omega^{TR} = \mathbf{E}_\omega^*(\mathbf{r}), \qquad \mathbf{H}_\omega(\mathbf{r}) \xrightarrow{\mathcal{T}} \mathbf{H}_\omega^{TR} = -\mathbf{H}_\omega^*(\mathbf{r}), \tag{3a}$$

$$\mathbf{j}_\omega(\mathbf{r}) \xrightarrow{\mathcal{T}} \mathbf{j}_\omega^{TR} = -\mathbf{j}_\omega^*(\mathbf{r}), \tag{3b}$$

The complex amplitudes of $\mathbf{D}$ and $\mathbf{B}$ are transformed similarly, taking into account the parity (even or odd) of each vector field.

In the lossless case, the response of a material is time-reversal invariant if and only if the material is reciprocal; for an explicit proof see Appendix A. In contrast, dissipative dielectrics are not time-reversal invariant, even though they are reciprocal.

## 3. Hidden symmetry

Realistic materials are lossy from an electromagnetic point of view: as a wave propagates in a medium part of its energy is irreversibly lost in the form of heat. The absorption effect is included in the dielectric response. Due to this reason, the equations of macroscopic electrodynamics are not time-reversal invariant when the materials are lossy. Nevertheless, as is well known, the response of dissipative systems is constrained by reciprocity relations [2, 6].

As mentioned in the Introduction, the equations that rule the propagation of light in natural media are time-reversal invariant at the microscopic level [2, 6]. The time-reversal of some wave process in a dissipative system requires the medium to pump the time-reversed field distribution, to give back all the energy originally dissipated as heat. No matter how strange as this may look, the theory predicts that if the system could be prepared in a suitable microscopic initial state it could be possible to generate the time-reversed wave dynamics, starting with an apparently chaotic thermal bath [1]. In practice, the preparation of the required initial state is unrealistic. The microscopic reversibility is a central argument of Onsager's theory [2, 6].

Next, I show –without invoking any microscopic arguments– that a reciprocal dissipative macroscopic system has a hidden time-reversal symmetry.

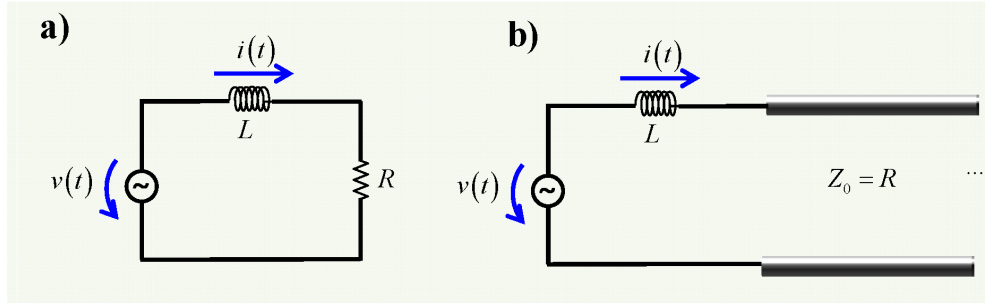

Fig. 1. a) Dissipative RL circuit. b) Equivalent time-reversal invariant lossless circuit where the resistor is implemented with a semi-infinite transmission line.

### 3.1 Dissipation implemented with an "open" system

Consider the circuit of Fig. 1a, which is formed by a resistor and an inductance fed by a voltage generator. The current circulating in the circuit is the solution of the differential equation:

$$Ri(t) + L\frac{di(t)}{dt} = v(t). \tag{4}$$

The voltage and the current are transformed under a time-reversal transformation as:

$$v(t) \xrightarrow{T} v(-t), \qquad i(t) \xrightarrow{T} -i(-t). \tag{5}$$

Thus, the system is time reversal invariant only when $R = 0$. The transfer function of the system in the frequency domain is $I(\omega)/V(\omega) = 1/(R - i\omega L)$, and as expected has a pole in the lower-half frequency plane due to the dissipation in the resistor.

From the theory of transmission lines, a semi-infinite lossless transmission line with characteristic impedance $Z_0$ is equivalent to a resistor $R = Z_0$. Therefore, the dissipative circuit of Fig. 1a has exactly the same response as the lossless circuit of Fig. 1b. This property was previously discussed in the context of the analysis of dissipative quantum systems [11]. The circuit of Fig. 1b is manifestly time-reversal invariant. Thus, it follows that an open time-reversal invariant system can mimic perfectly the response of a dissipative system. One may say that the circuit of Fig. 1a has a *hidden time-reversal symmetry*.

Likewise, the response of a lossy dielectric may be perfectly reproduced by an open time-reversal invariant lossless system. To show this, I consider without loss of generality that the dielectric response is determined by the Lorentz model:

$$\varepsilon(\omega) = 1 + \frac{\Omega_0^2}{\omega_0^2 - \omega^2 - i\Gamma\omega}. \tag{6}$$

The polarization vector **P** of a material with the Lorentz dispersion is described by the differential equation:

$$\frac{\partial^2 \mathbf{P}}{\partial t^2} + \Gamma \frac{\partial \mathbf{P}}{\partial t} + \omega_0^2 \mathbf{P} = \varepsilon_0 \Omega_0^2 \mathbf{E}. \tag{7}$$

The polarization vector **P** is even under a time-reversal: $\mathbf{P}(\mathbf{r},t) \xrightarrow{\mathcal{T}} \mathbf{P}(\mathbf{r},-t)$. Thus, the dynamics determined by (7) is not time-reversal invariant when the material is lossy ($\Gamma > 0$).

Similar to the circuit of Fig. 1a, it is possible to find a lossless open system with the same response as that of the lossy material (6). Here is an example inspired by transmission line theory:

$$\frac{\partial^2 \mathbf{P}}{\partial t^2} + \frac{2\Gamma}{AZ_0} \mathbf{V}_{u=0} + \omega_0^2 \mathbf{P} = \varepsilon_0 \Omega_0^2 \mathbf{E}, \tag{8a}$$

$$\frac{\partial \mathbf{V}_u}{\partial u} = -L \frac{\partial \mathbf{I}_u}{\partial t}, \qquad \frac{\partial \mathbf{I}_u}{\partial u} = -C \frac{\partial \mathbf{V}_u}{\partial t} + A \frac{\partial \mathbf{P}}{\partial t} \delta(u). \tag{8b}$$

I introduced two vector fields $\mathbf{V}_u, \mathbf{I}_u$ that depend on the spacetime coordinates $(\mathbf{r},t)$ and on a fictitious 4$^{th}$ space-coordinate $u$. The fields $\mathbf{P}, \mathbf{E}$ depend only on $(\mathbf{r},t)$. The vector field $\mathbf{V}_u = (V_u^x, V_u^y, V_u^z)$ has the units of a voltage and the field $\mathbf{I}_u = (I_u^x, I_u^y, I_u^z)$ has the units of a current. The parameters $L, C$ play the roles of a distributed inductance and capacitance, respectively, and $Z_0 = \sqrt{L/C}$ is the characteristic impedance. The parameter $A$ has units of area so that $A\partial \mathbf{P}/\partial t$ has units of current. The polarization vector is coupled to the Maxwell equations in the usual way:

$$\nabla \times \mathbf{E} = -\mu_0 \partial \mathbf{H}/\partial t, \qquad \nabla \times \mathbf{H} = \mathbf{j} + \partial \mathbf{P}/\partial t + \varepsilon_0 \partial \mathbf{E}/\partial t. \tag{8c}$$

The system (8) models a three-dimensional system (the material) coupled to a 4$^{th}$ dimension through a distributed network of 1D transmission lines. The transmission lines are infinitely extended along the 4$^{th}$ dimension, $-\infty < u < \infty$. The coupling between the material and the lines is determined by the voltages calculated at the point $u = 0$. The transmission line associated with $V_u^i, I_u^i$ ($i=x,y,z$) is fed by a current generator placed at $u = 0$ with amplitude $A\partial P_i/\partial t$. From (8b) it is clear that $I_{u=0^+}^i - I_{u=0^-}^i = A\partial P_i/\partial t$ and $V_{u=0^+}^i = V_{u=0^-}^i \equiv V_{u=0}^i$. If one imposes that $V_u^i, I_u^i$ satisfy radiation boundary conditions so that the energy can only flow away from the generator, it follows that $V_u^i/I_u^i = Z_0$ for $u > 0$ and $V_u^i/I_u^i = -Z_0$ for $u < 0$. Thus, in these conditions, $2V_{u=0}^i/Z_0 = A\partial P_i/\partial t$. Substituting this result into (8a) we recover the dissipative Lorentz model (7). Therefore, the dynamics of (7) is exactly reproduced by the lossless open system described by (8a)-(8b) when radiation boundary conditions are imposed on the transmission lines.

Moreover, the system (8) is time-reversal invariant because the dynamics of the time-reversed fields [$V_u^i, I_u^i$ are transformed as in (5)] is described by the same equations [Eq. (8)]. Thus, the dissipative Lorentzian response (6) has a hidden time reversal symmetry, consistent with the microscopic reversibility discussed by Onsager [2, 6].

The enunciated results can be readily extended to other more complex dispersive models (and to the anisotropic case), because the permittivity of a generic dispersive material can be written as a sum of Lorentz poles (see Appendix B).

*3.2 Dissipation imitated by a lossless material*

Consider again the system of Fig. 1a. It is interesting to note that the response of the resistor may also be imitated by a line with finite length $l$ terminated in short-circuit, i.e., by a closed lossless system. Specifically, for an excitation that starts at $t$=0, the response of the resistor is perfectly mimicked by the line in the time interval $0 < t < 2T_p$ with $T_p = l/v_p$ the time of propagation from the beginning to the end of the line. Thus, as $l \to \infty$ the response of the line becomes coincident with that of the dissipative system for longer and longer time intervals. Note that the impedance of the short-circuited line is purely reactive. When $l = \infty$ the line is semi-infinite, the system is open, and its response mimics exactly that of the resistor.

Furthermore, for an excitation of the system with a time variation $e^{-i\omega t} = e^{\omega'' t} e^{-i\omega' t}$ with $\omega = \omega' + i\omega''$ in the upper-half frequency plane (UHP) ($\omega'' > 0$), the line terminated with a short circuit may reproduce well the resistor response for any time instant. The reason is that when $l$ is sufficiently large the echoes coming from the transmission line have negligible amplitude as compared to the signal coming from the generator, because the amplitude of the latter grows exponentially with time ($e^{+\omega'' t}$ with $\omega'' > 0$). Thus the UHP response of a dissipative circuit element (the resistor) can be approximated as much as desired by that of a lossless reactive circuit (line terminated with a short circuit).

It is evident from the transmission-line model developed in Sect. 3.1 that the response of a lossy dielectric can be imitated during an arbitrarily large time period by the response of some lossless material. Furthermore, the UHP response of a lossy dielectric can be approximated as much as desired by that of a lossless time-reversal invariant material. An explicit proof of this result (not based on the transmission-line model of Sect. 3.1) is given in Appendix B.

In quantum theory it is often useful to model dissipation effects through the interaction of some Hermitian system with a bath of harmonic oscillators [7-12]. The formalism of Appendix B connects closely with such an idea, as it unveils that the dielectric function of a dissipative material is determined by a "bath" (continuum) of lossless Lorentz-type oscillators.

**4. Reciprocity theorem**

As discussed in Sect. 2, the concept of reciprocity is broader than the concept of time-reversal invariance. This property might suggest that reciprocity is rooted on some special characteristic (e.g., on the mathematical structure) of the Maxwell equations, beyond the time-reversal invariance. Next, I prove that the Lorentz reciprocity is essentially a consequence of the linearity and of the hidden time-reversal invariance of the materials.

*4.1 Lossless systems*

To begin with, I focus on lossless time-reversal invariant systems, and I derive the Lorentz reciprocity theorem without invoking directly the Maxwell's equations or the constitutive relations of the materials. The proof uses simply the conservation of energy and the "linearity" of the equations.

To this end, consider a generic solution $\mathbf{E}_\omega, \mathbf{H}_\omega$ of the Maxwell's equations in time-harmonic regime (with $\omega$ real-valued) created by some external current density $\mathbf{j}_\omega$. Assuming that the system is lossless, the conservation of energy (Poynting theorem) implies that $\nabla \cdot \mathbf{S}_{\text{av}} = p_{\text{ext,av}}$, where $\mathbf{S}_{\text{av}} = (1/2)\text{Re}\{\mathbf{E}_\omega \times \mathbf{H}_\omega^*\}$ is the time-averaged energy-density flux

(Poynting vector) and $p_{\text{ext,av}} = -(1/2)\text{Re}\{\mathbf{E}_\omega \cdot \mathbf{j}_\omega^*\}$ is the time-averaged power extracted from the external current $\mathbf{j}_\omega$ per unit of volume. The energy balance constraint can be expressed as

$$\text{Re}\{\nabla \cdot [\mathcal{S}(\mathbf{E}_\omega, \mathbf{H}_\omega^*)] - \mathcal{P}_{\text{ext}}(\mathbf{E}_\omega, \mathbf{j}_\omega^*)\} = 0, \tag{9}$$

where for convenience I introduced the bilinear forms $\mathcal{S}(\mathbf{E}, \mathbf{H}) \equiv \mathbf{E} \times \mathbf{H}$ and $\mathcal{P}_{\text{ext}}(\mathbf{E}, \mathbf{j}) \equiv -\mathbf{E} \cdot \mathbf{j}$. The bilinear forms satisfy $[\mathcal{S}(\mathbf{E}, \mathbf{H})]^* = \mathcal{S}(\mathbf{E}^*, \mathbf{H}^*)$ and $[\mathcal{P}_{\text{ext}}(\mathbf{E}, \mathbf{j})]^* = \mathcal{P}_{\text{ext}}(\mathbf{E}^*, \mathbf{j}^*)$.

Let $\mathbf{E}'_\omega, \mathbf{H}'_\omega$ and $\mathbf{E}''_\omega, \mathbf{H}''_\omega$ be field distributions created respectively by $\mathbf{j}'_\omega$ and $\mathbf{j}''_\omega$. Then, by linearity $\mathbf{E}'_\omega + \mathbf{E}''_\omega, \mathbf{H}'_\omega + \mathbf{H}''_\omega$ is the solution of the Maxwell equations for the excitation $\mathbf{j}'_\omega + \mathbf{j}''_\omega$, and thus it must satisfy (9). This is only possible if:

$$\text{Re}\{\nabla \cdot [\mathcal{S}(\mathbf{E}'_\omega, \mathbf{H}''^*_\omega) + \mathcal{S}(\mathbf{E}''^*_\omega, \mathbf{H}'_\omega)] - \mathcal{P}_{\text{ext}}(\mathbf{E}'_\omega, \mathbf{j}''^*_\omega) - \mathcal{P}_{\text{ext}}(\mathbf{E}''^*_\omega, \mathbf{j}'_\omega)\} = 0. \tag{10}$$

Now, if the system has the time-reversal symmetry $\mathbf{E}''^*_\omega, -\mathbf{H}''^*_\omega, -\mathbf{j}''^*_\omega$ is also a solution of the Maxwell equations in the same physical platform. Thereby, it is possible to replace $(\mathbf{E}''_\omega, \mathbf{H}''_\omega, \mathbf{j}''_\omega)$ by $e^{i\phi}(\mathbf{E}''^*_\omega, -\mathbf{H}''^*_\omega, -\mathbf{j}''^*_\omega)$ in Eq. (10), and the resulting equation remains true for arbitrary $\mathbf{E}'_\omega, \mathbf{H}'_\omega$ and $\mathbf{E}''_\omega, \mathbf{H}''_\omega$. The constant factor $e^{i\phi}$ is introduced by convenience (this is possible due to the linearity of the system). Thus, it follows that for any $\phi$:

$$\text{Re}\{e^{-i\phi}[\nabla \cdot [-\mathcal{S}(\mathbf{E}'_\omega, \mathbf{H}''_\omega) + \mathcal{S}(\mathbf{E}''_\omega, \mathbf{H}'_\omega)] + \mathcal{P}_{\text{ext}}(\mathbf{E}'_\omega, \mathbf{j}''_\omega) - \mathcal{P}_{\text{ext}}(\mathbf{E}''_\omega, \mathbf{j}'_\omega)]\} = 0. \tag{11}$$

If $w$ is a complex number and $\text{Re}\{e^{-i\phi}w\} = 0$ for an arbitrary $\phi$ then $w = 0$. Thereby, it follows that

$$\nabla \cdot [-\mathcal{S}(\mathbf{E}'_\omega, \mathbf{H}''_\omega) + \mathcal{S}(\mathbf{E}''_\omega, \mathbf{H}'_\omega)] = -\mathcal{P}_{\text{ext}}(\mathbf{E}'_\omega, \mathbf{j}''_\omega) + \mathcal{P}_{\text{ext}}(\mathbf{E}''_\omega, \mathbf{j}'_\omega), \tag{12}$$

which is exactly the Lorentz reciprocity theorem in the differential form.

The above analysis is rather general, and only requires that the system is linear, non-dissipative, time-reversal invariant, and that the energy density flux is determined by $\mathbf{E} \times \mathbf{H}$ (local media). Clearly, the derivation can be readily generalized to other physical systems (unrelated to electromagnetism) wherein the energy density flux and the extracted power density are bilinear forms of the relevant fields. This property helps to understand why the reciprocity constraint is so ubiquitous in "wave systems" [2, 6].

*4.2 Lossy systems*

The analysis of the previous subsection can be extended to general lossy reciprocal materials with a hidden-time reversal symmetry. For simplicity, the following discussion is focused on the model of Sect. 2.1.

For the system described by Eqs. (8), the conservation of energy in time-harmonic regime is expressed by $\nabla \cdot \mathbf{S}_{\text{av}} = p_{\text{ext,av}} - p_{\text{d,av}}$ with $p_{\text{d,av}} = (\Upsilon/2)\text{Re}\{\mathbf{V}_{u=0,\omega} \cdot \delta\mathbf{I}^*_{\omega,u=0}\}$, $\Upsilon = 2\Gamma/(\varepsilon_0\Omega_0^2 Z_0 A^2)$, and $\delta\mathbf{I}_{\omega,u=0} = \mathbf{I}_{u=0^+,\omega} - \mathbf{I}_{u=0^-,\omega}$. Proceeding as in Sect. 4.1 and using the time-reversal invariance of the system (8), it is possible to show that:

$$\begin{aligned}\nabla \cdot [-\mathcal{S}(\mathbf{E}'_\omega, \mathbf{H}''_\omega) + \mathcal{S}(\mathbf{E}''_\omega, \mathbf{H}'_\omega)] &= -\mathcal{P}_{\text{ext}}(\mathbf{E}'_\omega, \mathbf{j}''_\omega) + \mathcal{P}_{\text{ext}}(\mathbf{E}''_\omega, \mathbf{j}'_\omega) \\ &\quad + \mathcal{P}_{\text{d}}(\mathbf{V}'_{\omega,u=0}, \delta\mathbf{I}''_{\omega,u=0}) - \mathcal{P}_{\text{d}}(\mathbf{V}''_{\omega,u=0}, \delta\mathbf{I}'_{\omega,u=0})\end{aligned} \tag{13}$$

with $\mathcal{P}_\mathrm{d}(\mathbf{V},\mathbf{I}) = \Upsilon \mathbf{V} \cdot \mathbf{I}$. As seen in Sect. 4.1, when radiation boundary conditions are enforced on the transmission lines the corresponding voltages and currents are linked as $\mathbf{V}_{u=0,\omega} = (Z_0/2)\delta\mathbf{I}_{u=0,\omega}$. In these conditions, the term $\mathcal{P}_\mathrm{d}(\mathbf{V}'_{\omega,u=0},\delta\mathbf{I}''_{\omega,u=0}) - \mathcal{P}_\mathrm{d}(\mathbf{V}''_{\omega,u=0},\delta\mathbf{I}'_{\omega,u=0})$ vanishes, and Eq. (13) yields the standard reciprocity theorem in a dissipative material system. Therefore, the reciprocity of dissipative systems is deeply rooted on the hidden time reversal invariance.

## 5. Summary

The response of an idealized "open" lossless time-reversal invariant system can be made exactly coincident with that of a lossy dielectric. Thus, dissipative (reciprocal) dielectrics have a hidden time-reversal symmetry and their response is fundamentally constrained by time-reversal invariance. In particular, it was shown that the reciprocity of dissipative systems is ultimately a consequence of the hidden time-reversal symmetry and of the linearity of the materials.

Furthermore, it was demonstrated that the dynamics of a lossy material can be precisely mimicked by a lossless material during an arbitrarily large time interval. In addition, the UHP response of a lossy material can be approximated as much as desired by that of some lossless material, without any restriction on the duration of the excitation.

## Appendix A

Suppose that the complex amplitudes of the electromagnetic fields are related by bianisotropic constitutive relations of the type,

$$\begin{pmatrix} \mathbf{D}_\omega \\ \mathbf{B}_\omega \end{pmatrix} = \underbrace{\begin{pmatrix} \varepsilon_0 \overline{\varepsilon} & \frac{1}{c}\overline{\xi} \\ \frac{1}{c}\overline{\zeta} & \mu_0 \overline{\mu} \end{pmatrix}}_{\mathbf{M}(\omega)} \begin{pmatrix} \mathbf{E}_\omega \\ \mathbf{H}_\omega \end{pmatrix}. \tag{A1}$$

The material matrix $\mathbf{M}(\omega)$ is written in terms of permittivity, permeability, and magneto-electric coupling tensors, in a standard way. Then, the time-reversed fields $\mathbf{D}_\omega^{\mathrm{TR}}, \mathbf{B}_\omega^{\mathrm{TR}}$ and $\mathbf{E}_\omega^{\mathrm{TR}}, \mathbf{H}_\omega^{\mathrm{TR}}$ are linked by a material matrix $\mathbf{M}^{\mathrm{TR}}(\omega)$ given by:

$$\mathbf{M}^{\mathrm{TR}}(\omega) = \boldsymbol{\sigma}_z \cdot \mathbf{M}^*(\omega) \cdot \boldsymbol{\sigma}_z, \quad \text{with} \quad \boldsymbol{\sigma}_z = \begin{pmatrix} \mathbf{1}_{3\times 3} & \mathbf{0} \\ \mathbf{0} & -\mathbf{1}_{3\times 3} \end{pmatrix}. \tag{A2}$$

Note that because of the reality condition, $\mathbf{M}^*(\omega) = \mathbf{M}(-\omega^*)$, the above formula is equivalent to $\mathbf{M}^{\mathrm{TR}}(\omega) = \boldsymbol{\sigma}_z \cdot \mathbf{M}(-\omega) \cdot \boldsymbol{\sigma}_z$.

A system formed by bianisotropic materials is time-reversal invariant when $\mathbf{M}^{\mathrm{TR}}(\omega) = \mathbf{M}(\omega)$. From Eq. (A2), the general conditions for time-reversal invariance are $\overline{\varepsilon} = \overline{\varepsilon}^*$, $\overline{\mu} = \overline{\mu}^*$, and $\overline{\xi} = -\overline{\xi}^*$ and $\overline{\zeta} = -\overline{\zeta}^*$. For example, dissipative isotropic dielectrics are not time reversal invariant because $\varepsilon \neq \varepsilon^*$.

For a lossless medium, the material matrix is necessarily Hermitian symmetric, $\mathbf{M}(\omega) = \mathbf{M}^\dagger(\omega)$, for $\omega$ real-valued [22]. In these conditions, Eq. (A2) reduces to (the superscript $T$ stands for the matrix transpose):

$$\mathbf{M}^{\mathrm{TR}}(\omega) = \boldsymbol{\sigma}_z \cdot \mathbf{M}^T(\omega) \cdot \boldsymbol{\sigma}_z, \qquad \text{for } \omega \text{ real-valued.} \tag{A3}$$

This constraint is equivalent to the well-known Lorentz reciprocity relations: $\bar{\bar{\varepsilon}} = \bar{\bar{\varepsilon}}^T$, $\bar{\bar{\mu}} = \bar{\bar{\mu}}^T$, and $\bar{\bar{\xi}} = -\bar{\bar{\zeta}}^T$. Thus, lossless time-reversal invariant linear materials are always reciprocal.

**Appendix B**

In this Appendix, I show that a generic *passive* reciprocal medium can be regarded as the limit of some sequence of time-reversal invariant lossless materials. For simplicity, I restrict the analysis to isotropic dissipative dielectrics described by some dispersive permittivity $\varepsilon = \varepsilon(\omega)$. The permittivity is assumed to be a meromorphic function of frequency and hence has a partial-fraction expansion of the form:

$$\varepsilon(\omega) = 1 + \sum_n \frac{-\Omega_n}{\omega - \omega_{p,n}} = 1 + \sum_{\text{Re}\{\omega_{p,n}\}>0} \left( \frac{-\Omega_n}{\omega - \omega_{p,n}} + \frac{\Omega_n^*}{\omega + \omega_{p,n}^*} \right). \tag{B1}$$

Here, $\omega_{p,n}$ are the poles and $-\Omega_n$ are the residues of the permittivity. The second identity uses the reality-condition $\varepsilon(\omega) = [\varepsilon(-\omega^*)]^*$. The passivity of the material requires that the poles are in the lower-half frequency plane (for a time variation $e^{-i\omega t}$), i.e., $\gamma_n = -\text{Im}\{\omega_{p,n}\} > 0$. Furthermore, in order that $\text{Im}\{\varepsilon\} > 0$ in the positive real axis, it is necessary that $\varepsilon(\omega)$ decays as $1/\omega^2$ in the upper-half of the imaginary frequency axis [22]. This is only possible if $\Omega_n$ is a positive real-number. Thus, writing $\omega_{p,n} = \omega_n - i\gamma_n$, it follows that the permittivity of a passive material is necessarily of the form:

$$\varepsilon(\omega) = 1 - \sum_n \Omega_n \left( \frac{1}{\omega - \omega_n + i\gamma_n} + \frac{-1}{\omega + \omega_n + i\gamma_n} \right). \tag{B2}$$

In the above, $\omega_n, \gamma_n$ determine the resonant frequency and damping rate, respectively, and are positive numbers. As expected, each term of the sum is equivalent to a Lorentz oscillator [16]. For example, when the summation restricted to the $n=1$ term, one obtains the Lorentzian response (6) with $\Gamma = 2\gamma_1$, $\omega_0 = \sqrt{\omega_1^2 + \gamma_1^2}$, and $\Omega_0 = \sqrt{2\Omega_1 \omega_1}$.

Now the key observation is that in the upper-half frequency plane (UHP) ($\text{Im}\{\omega\} > 0$) one has:

$$\frac{1}{\omega - \omega_n + i\gamma_n} = \frac{1}{\pi} \int_{-\infty}^{+\infty} d\xi \frac{\gamma_n}{\gamma_n^2 + (\xi - \omega_n)^2} \frac{1}{\omega - \xi}. \tag{B3}$$

The identity can be verified using the Cauchy theorem. Therefore, the dielectric function in the UHP may be written as:

$$\varepsilon(\omega) = 1 + \int_{-\infty}^{+\infty} d\xi \, \chi_\xi(\omega), \quad \chi_\xi(\omega) = -\frac{1}{\pi} \left( \frac{1}{\omega - \xi} - \frac{1}{\omega + \xi} \right) \sum_n \frac{\Omega_n \gamma_n}{\gamma_n^2 + (\xi - \omega_n)^2}. \tag{B4}$$

Suppose that the integral is approximated by a discrete summation of the form $\varepsilon(\omega) \approx \varepsilon_{\Delta\xi}(\omega)$, with $\varepsilon_{\Delta\xi}(\omega) = 1 + \sum_{m=-M}^{M} \Delta\xi \times \chi_{m\Delta\xi}(\omega)$. The approximation is obtained by sampling $2M+1$ points of the integration domain spaced by $\Delta\xi$. Evidently, one has $\varepsilon(\omega) = \lim_{\substack{\Delta\xi \to 0^+ \\ M\Delta\xi \to \infty}} \varepsilon_{\Delta\xi}(\omega)$ when $\omega$ is in the UHP. The function $\varepsilon_{\Delta\xi}(\omega)$ is analytic in the complex plane, with the exception of the real-frequency axis where $\varepsilon_{\Delta\xi}$ has poles at $\omega = m\Delta\xi$, $m = \pm 1, \pm 2, \ldots$. Since for $\omega$ real-valued $d\left[\omega\varepsilon_{\Delta\xi}(\omega)\right]/d\omega > 0$, the function $\varepsilon_{\Delta\xi}$ describes the response of a lossless passive system [22]. In addition, the material response is time-reversal invariant due to $\varepsilon_{\Delta\xi}(\omega) = \varepsilon_{\Delta\xi}(-\omega)$.

In any compact region of the UHP, $\varepsilon(\omega)$ can be approximated as much as desired by some $\varepsilon_{\Delta\xi}(\omega)$ with the sampling period small enough and *M* large enough. Thus, it follows that a lossy reciprocal dielectric may always be regarded as the limit of a sequence of time-reversal invariant lossless materials. This further supports that the reciprocal properties of a lossy dielectric are inherited from the hidden time-reversal symmetry.

Even though the previous discussion was focused on isotropic dielectrics, the results can be generalized to other reciprocal and passive material platforms.

## Acknowledgments

This research was funded by the IET under the A F Harvey Engineering Research Prize and by Fundação para Ciência e a Tecnologia (FCT) under projects PTDC/EEITEL/4543/2014 and UID/EEA/50008/2017. The author gratefully acknowledges stimulating discussions with Dimitrios Tzarouchis.